\documentclass[twocolumn,epjc3]{svjour3}  
\RequirePackage{graphicx}
\RequirePackage[colorlinks,citecolor=blue,urlcolor=blue,linkcolor=blue]{hyperref}
\journalname{Eur. Phys. J. C}

\usepackage{lineno,hyperref}
\usepackage{multirow}
\usepackage{comment}
\usepackage{amssymb}
\usepackage{amsmath}
\usepackage{graphicx,color}
\usepackage{cite}
\usepackage{xspace}	

\newcommand{\roots} {\mbox{$\sqrt{s_{\rm NN}}$}}

\newcommand{\PYTHIA}{\textsc{Pythia}8}
\newcommand{\PYTHIAV}{\textsc{Pythia}8.3}
\newcommand{\pt}    {$p_{\mathrm{T}}$}
\newcommand{\dq}    {$\Delta Q$}
\newcommand{\dy}    {$\Delta y$}
\begin{document}

\title{Measuring charge transport in high-energy nuclear collisions with an energy scan of isobaric collisions}

\author{Wendi Lv\thanksref{addr1, e1}
        \and
        Niseem Magdy\thanksref{addr2, addr3, e2}
        \and
        Rongrong Ma\thanksref{addr3, e3}
        \and 
        Zebo Tang \thanksref{addr1, e4}
        \and
        Prithwish Tribedy \thanksref{addr3, e5}
        \and
        Chun Yuen Tsang
        \thanksref{addr4, e6}
        \and
        Zhangbu Xu \thanksref{addr3,addr4, e7}
}

\thankstext{e1}{wdlv@mail.ustc.edu.cn}
\thankstext{e2}{niseem.abdelrahman@tsu.edu}
\thankstext{e3}{marr@bnl.gov}
\thankstext{e4}{zbtang@ustc.edu.cn}
\thankstext{e5}{ptribedy@bnl.gov}
\thankstext{e6}{ctsang@bnl.gov}
\thankstext{e7}{zxu22@kent.edu}

\institute{Department of Modern Physics, University of Science and Technology of China, Hefei, Anhui 230026, China \label{addr1}
        \and
          Department of Physics, Texas Southern University, Houston, TX 77004, USA\label{addr2}
          \and
          Department of Physics, Brookhaven National Laboratory, Upton, NY 11973, USA\label{addr3}
          \and
          Department of Physics, Kent State University, Kent, OH 44242, USA\label{addr4}
}

\date{Received: date / Accepted: date}

\maketitle


\begin{abstract}
We present a method to measure electric-charge transport in high-energy nuclear collisions using a beam-energy scan of isobaric systems. Comparing collisions of nuclei with identical mass number but different atomic number allows the charge difference ($\Delta Q$) to be extracted with a double-ratio technique that suppresses most experimental systematic uncertainties. By varying the beam energy, the rapidity gap ($\Delta y$) over which electric charge is transported can be systematically scanned. Simulations of Ru+Ru and Zr+Zr collisions at $\sqrt{s_{\rm NN}}=19.6$--200~GeV with UrQMD and \PYTHIA\ Angantyr show that midrapidity $\Delta Q$ decreases exponentially with increasing $\Delta y$, with the slope parameter exhibiting strong model dependence. Comparisons to baryon-number transport reveal distinct patterns. In both UrQMD and \PYTHIA\ Angantyr (with and without final-state baryon junctions), where baryon number is carried solely by valence quarks, the rapidity slope for baryon transport is larger than that for electric-charge transport. In contrast, scenarios that include baryon junctions in the initial state are expected to produce the opposite trend. This demonstrates that an isobar beam-energy scan provides a sensitive probe of electric-charge transport and offers new constraints on the microscopic mechanisms governing conserved-charge redistribution in QCD matter.
\end{abstract}


\section{INTRODUCTION}

A central goal of ultra-relativistic heavy-ion collision experiments is to map the phase diagram of Quantum Chromodynamics (QCD) and characterize the properties of the Quark–Gluon Plasma (QGP)~\cite{STAR:2005gfr,Shuryak:2024zrh, Harris:2023tti}. In such collisions, conserved quantities, including baryon number ($B$) and electric charge ($Q$), are initially carried by the incoming nuclei and subsequently redistributed throughout the hot and dense medium. Understanding how these conserved quantities are transported away from the beam rapidity is essential for constraining both the initial conditions of QGP formation and its evolution.

While baryon-number transport has been extensively measured at RHIC and the LHC~\cite{BRAHMS:2003wwg, BRAHMS:2009wlg, STAR:2008med, STAR:2017sal, ALICE:2010hjm, ALICE:2013mez,STAR:2024lvy,Tsang:2024zsq}, direct measurements of electric-charge transport remain scarce. This is largely because the produced medium is nearly charge symmetric around midrapidity, making net-charge signals small compared to the large background from particle–antiparticle pair production. On the other hand, it is well established that electric charge is carried by valence quarks, whereas the microscopic carrier of baryon number remains under debate. Proposed mechanisms include valence-quark transport as well as Y-shaped gluonic junctions~\cite{Artru:1974zn,Rossi:1977cy,Kharzeev:1996sq,Suganuma:2004zx,Magdy:2025udq,Magdy:2024dpm,STAR:2024lvy}. Therefore, measurements of electric-charge transport provide a useful reference for testing whether baryon number follows valence-quark transport or requires additional transport mechanisms.

The electric charge of a system can be quantified by its net electric charge, defined as
\begin{equation}
Q = N_{+}-N_{-},
\label{eq:Q}
\end{equation}
where $N_{+}$ and $N_{-}$ denote the total numbers of positively and negatively charged particles, respectively. Since $\pi$, $K$, and $p$ dominate the charged hadron yields, $Q$ may also be written as
\begin{equation}
Q=(N_{\pi^{+}}+N_{K^{+}}+N_{p})-(N_{\pi^{-}}+N_{K^{-}}+N_{\bar{p}}).
\label{eq:net-Q}
\end{equation}
Because pions dominate the charged multiplicity and $N_{\pi^{+}} \approx N_{\pi^{-}}$ with large absolute values, extracting $Q$ with high precision is challenging.

A recent proposal~\cite{Lewis:2022arg} demonstrated that the charge difference (\dq) between two isobar systems—nuclei with identical mass number $A$ but different atomic number $Z$—can be determined precisely using a double-ratio method. Defining
\begin{equation}
\Delta Q = Q_{\rm Iso1} - Q_{\rm Iso2},
\label{eq:dQ_exact}
\end{equation}
and exploiting the near equality of particle multiplicities in the two systems, the charge difference can be approximated as
\begin{equation}
\Delta Q \approx N_{\pi}(R2_{\pi}-1) + N_{K}(R2_{K}-1) + N_{p}(R2_{p}-1),
\label{eq:dQ}
\end{equation}
where $N_{\pi}$, $N_{K}$, and $N_{p}$ are the average yields and the $R2$ factors represent the corresponding double ratios:
\begin{equation}
R2_{\pi} = 
\frac{(N_{\pi^{+}}/N_{\pi^{-}})_{\rm Iso1}}
     {(N_{\pi^{+}}/N_{\pi^{-}})_{\rm Iso2}}.
\end{equation}
The double-ratio technique suppresses nearly all experimental systematic uncertainties, enabling a precise measurement of \dq. This method has been successfully applied to analyze the isobar collisions at RHIC top energy~\cite{STAR:2024lvy}.

Electric-charge transport can then be characterized by studying \dq\ as a function of the rapidity gap
\begin{equation}
\Delta y = y_{\rm beam} - y_{Q},
\end{equation}
where $y_{\rm beam}$ is the beam rapidity and $y_{Q}$ is the rapidity at which \dq\ is measured. In symmetric collisions, both beams contribute to \dq\ at a fixed $y_{Q}$, and the limited rapidity coverage of modern detectors restricts the accessible $\Delta y$ range. A beam-energy scan (BES) overcomes these limitations: by varying $y_{\rm beam}$ while keeping $y_{Q}$ fixed, the BES effectively scans a wide range of $\Delta y$ with well-defined charge transport ($\Delta Q$). 

In this work, we demonstrate this method using simulations of $_{44}^{96}\mathrm{Ru}+{}_{44}^{96}\mathrm{Ru}$ and $_{40}^{96}\mathrm{Zr}+{}_{40}^{96}\mathrm{Zr}$ collisions at $\sqrt{s_{\rm NN}}=$ 200, 62.4, 39, 27, and 19.6~GeV, corresponding to a span of approximately 2.5 units in beam rapidity. We employ two event generators: Ultra-relativistic Quantum Molecular Dynamics (UrQMD)~\cite{Bass:1998ca, Bleicher:1999xi} and \PYTHIAV\ Angantyr~\cite{Bierlich:2018xfw}. Although neither model is expected to provide an accurate description of the full dynamics of heavy-ion collisions, they serve as useful tools for illustrating the proposed methodology. We further compare electric-charge transport with baryon-number transport to gain insight into the underlying carriers of baryon number~\cite{STAR:2024lvy, Lewis:2022arg}.

\section{EVENT GENERATOR}\label{sec:2}

The UrQMD model is based on a microscopic transport approach in which hadrons propagate covariantly along classical trajectories and interact via stochastic binary scatterings and resonance decays~\cite{Bass:1998ca, Bleicher:1999xi}. At low energies, UrQMD describes nuclear collisions in terms of interactions among established hadrons and their resonances, while at high energies, particle production is dominated by color-string excitation followed by string fragmentation into hadrons \cite{Bleicher:1999xi}. This provides an event-by-event space--time evolution from an initial non-\hspace{0pt}equilibrium stage through hadronic scatterings to a final set of stable hadrons and decay products \cite{Bass:1998ca, Bleicher:1999xi}. For hadronic scatterings, UrQMD uses inelastic cross sections that are either tabulated, parameterized, or constructed from other channels using general principles such as detailed balance and the additive quark model. In particular, both the baryon number and the electric charge are carried solely by valence quarks in UrQMD, and therefore their transport properties are expected to be closely connected~\cite{Lv:2023fxk}.

\PYTHIA\ is a general-purpose generator for high-energy proton-proton and lepton–hadron collisions, implementing multi-parton interactions (MPI), parton showers, and Lund-string hadronization~\cite{Sjostrand:2014zea}. The Angantyr extension~\cite{Bierlich:2018xfw} simulates heavy-ion collisions as a superposition of nucleon–nucleon sub-collisions determined from Glauber geometry (via \textsc{GLISSANDO})~\cite{Glauber:1955qq, Rybczynski:2013yba} and classified as elastic, diffractive, or absorptive. It can reproduce global features, such as particle multiplicities, transverse momentum (\pt) spectra, in heavy-ion collisions without modeling the QGP effects~\cite{Bierlich:2018xfw}.

In \PYTHIA, electric charges are carried by valence quarks while baryon formation and transport are controlled by string topologies set by color flow (CF) and color reconnection (CR). Two configurations are used depending on whether dynamical formation of the gluonic junction, also called baryon junction (B-J), during hadronization plays a significant role:
\begin{itemize}
  \item \textbf{\PYTHIA\ (without B-J):\\}
  Leading-color ($N_c\!\to\!\infty$) CF with the standard MPI-based CR that reconnects dipoles by minimizing the total string length~\cite{Sjostrand:2004pf, Christiansen:2015yqa}. Baryons mainly arise from di-quark pair production during Lund fragmentation, and junction production is rare. Consequently, baryon number tends to follow beam remnants, with limited transport to midrapidity.
  \item \textbf{\PYTHIA\ (with B-J): \\} 
  Beyond leading-color CF with an approximate SU(3) treatment and advanced CR that permits dynamical formation of junctions, anti-junctions, and multi-parton color topologies~\cite{Lonnblad:2023stc}. Junction-enabled strings enhance baryon yields and allow efficient transport of baryon number over large rapidity intervals.
\end{itemize}
It is worth noting that, in both configurations of \PYTHIA, baryon junctions are neither present in the incoming nuclei nor involved in the scattering processes. Therefore, the \PYTHIA\ simulation with baryon junctions enabled can only be regarded as a phenomenological implementation of junction-like topologies during hadronization, rather than as a realization of genuine initial-state baryon-junction transport.
 
As $^{96}_{44}\text{Ru}$ and $^{96}_{40}\text{Zr}$ nuclei are not included in the default \PYTHIA\ database, they are implemented as user-defined states in this work. We adopt an approach in which both species are modeled as spherical nuclei defined solely by their numbers of protons and neutrons, without incorporating additional nuclear structure effects such as neutron skins or nuclear deformations. Although these nuclear structure effects are expected to differ between Ru and Zr nuclei~\cite{Hammelmann:2019vwd,Jia:2022qgl,Zhang:2021kxj} and may quantitatively affect the predicted charge number, they are not expected to qualitatively modify the energy dependence of charge transport, which is the primary focus of this work. This is because the underlying differences in the Ru and Zr density distributions are intrinsic properties of the nuclei and are therefore not expected to vary with beam energy. The implementation of realistic density profiles for Ru and Zr nuclei would be needed in future precision predictions.

The event centrality in \PYTHIA\ is defined using the distribution of the number of participating nucleons ($N_{\text{part}}$) calculated from the Glauber geometry~\cite{Miller:2007ri} for each collision. Standard centrality classes are constructed by applying percentile cuts on the $N_{\text{part}}$ distribution, enabling a systematic comparison of observables as a function of collision geometry in both the Ru+Ru and Zr+Zr systems. On the other hand, the UrQMD simulations define centrality directly through the impact parameter ($b$) of the collision. Since $b$ and $N_{\text{part}}$ are monotonically and strongly correlated through the underlying nuclear overlap geometry, i.e., smaller impact parameters correspond to larger $N_{\text{part}}$, the two centrality definitions are effectively equivalent. Therefore, the different implementations are not expected to affect the comparison of results between the two models.

\section{RESULTS}\label{sec:3}

\begin{figure*}[htb]
\centering{
\includegraphics[width=0.8\linewidth, angle=-0,keepaspectratio=true,clip=true]{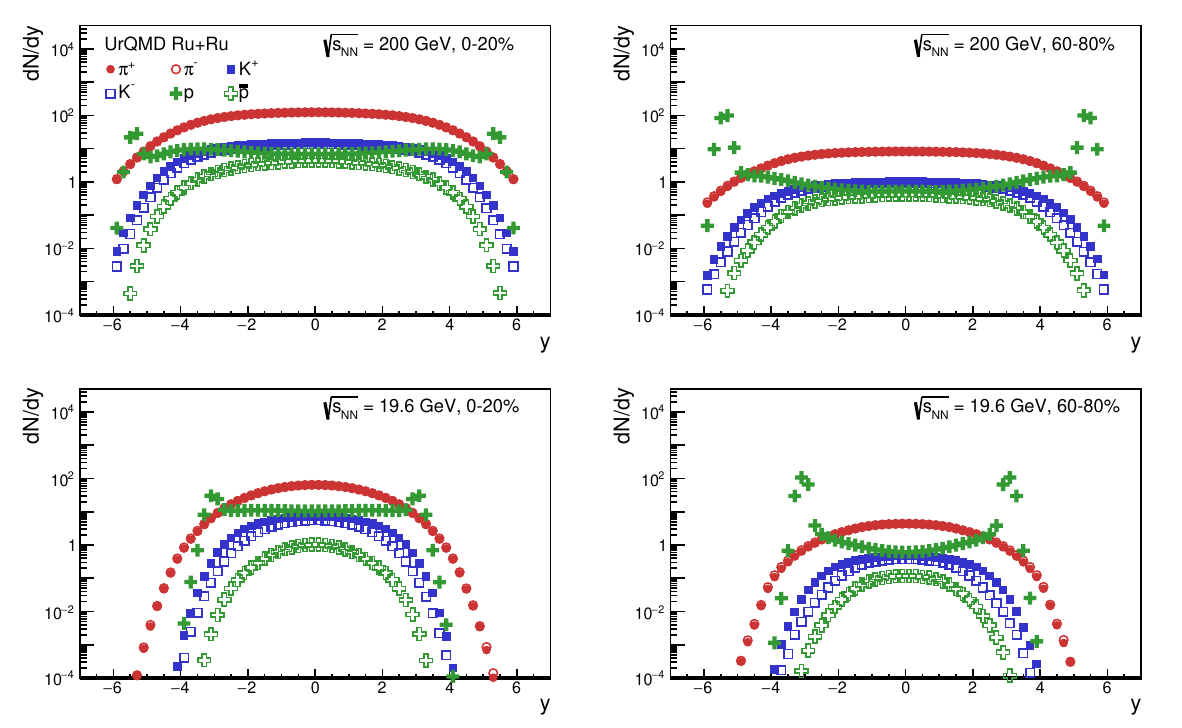}
\vskip -0.0cm
\caption{UrQMD simulations of rapidity distributions of $\pi^{\pm}$, $K^{\pm}$, $p$, and $\bar p$ for 0--20\% (left) and 60--80\% (right) centrality classes in Ru+Ru collisions at \roots\ = 200 (top) and 19.6 (bottom) GeV. 
\label{fig:fig1}
}
}
\vskip -0.0cm
\end{figure*}
Figure~\ref{fig:fig1} shows the rapidity distributions ($dN/dy$) of charged hadrons ($\pi^{\pm}, K^{\pm}, p, \bar{p}$) in 0--20\% central and 60--80\% peripheral Ru+Ru collisions at \roots\ = 200 and 19.6 GeV, simulated with the UrQMD model. As aforementioned, pions dominate the charged hadron yields around $y\sim0$, and the yields of $\pi^{+}$ and $\pi^{-}$ are nearly identical. On the other hand, protons are the most abundant charged hadrons near beam rapidity, which is 5.4 and 3.0 for 200 and 19.6 GeV collisions, respectively. As the collision energy decreases from 200 to 19.6 GeV, the rapidity distributions of all species become noticeably narrower, reflecting the reduced phase space available for particle production at lower center-of-mass energies. At the same time, the lower beam energy facilitates the transport of valence quarks, and thus electric charges, from beam rapidity toward midrapidity.

\begin{figure}[htb]
\centering{
\includegraphics[width=0.9\linewidth, angle=-0,keepaspectratio=true,clip=true]{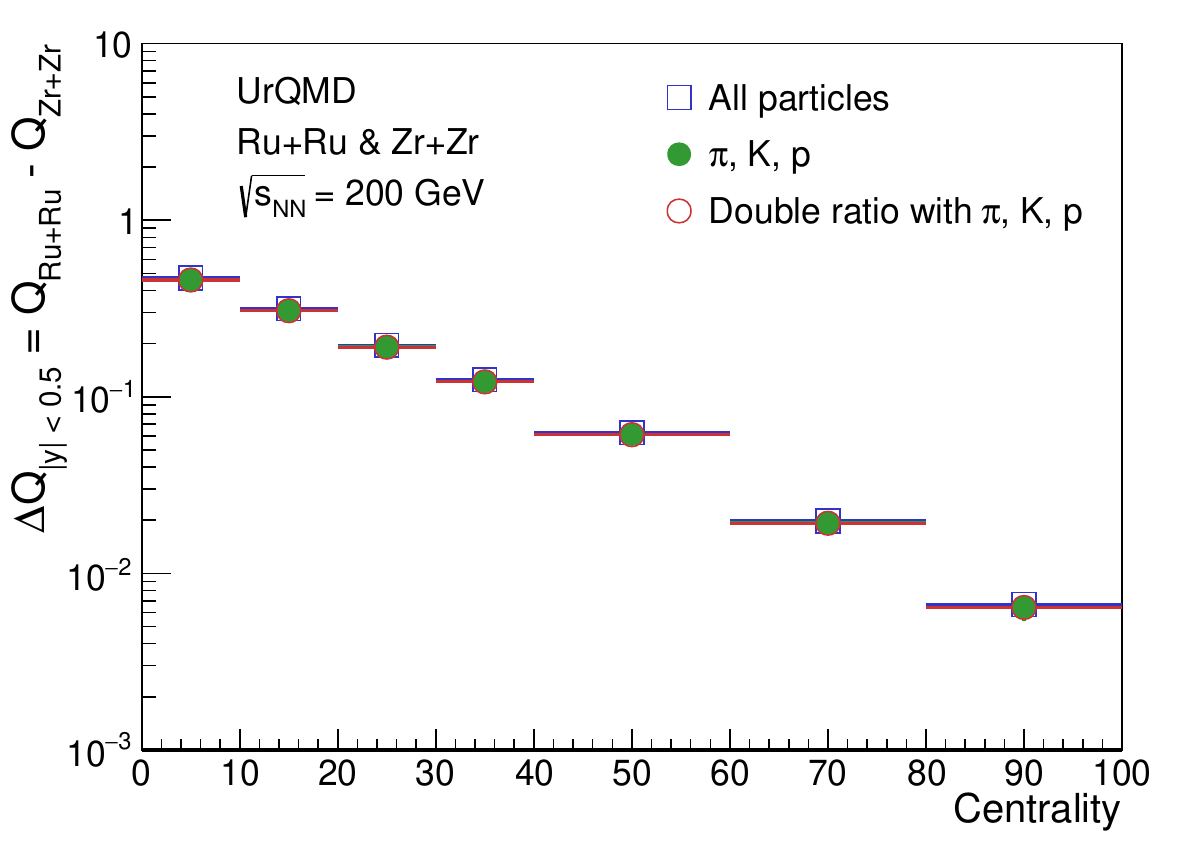}
\vskip -0.0cm
\caption{
Centrality dependence of $\Delta Q$ within $|y|<0.5$ for (i) all charged particles, (ii) charged $\pi$, $K$, $p$, and (iii) charged $\pi$, $K$, $p$ based on the double ratio method, for 200 GeV Ru+Ru and Zr+Zr collisions simulated with UrQMD.
\label{fig:fig2}
}
}
\vskip -0.0cm
\end{figure}

Using the $\pi$, $K$, $p$ yields shown in Fig.~\ref{fig:fig1}, $Q$ for Ru+Ru and Zr+Zr collisions within $|y|<0.5$ is calculated via Eq.~\ref{eq:net-Q}, and the corresponding \dq\ is obtained using Eq.~\ref{eq:dQ_exact}. The results, shown as filled circles in Fig.~\ref{fig:fig2}, exhibit an increase in \dq\ from peripheral to central collisions, reflecting the enhanced transport of electric charge to midrapidity resulting from the larger number of nucleon--nucleon interactions in central events. 
In simulations, where charged hadron yields are fully accessible, Eq.~\ref{eq:net-Q} can be applied directly to determine $Q$. In experimental measurements, however, such a procedure would suffer from large systematic uncertainties due to finite detector acceptance and efficiency. Instead, the double ratio method~\cite{Lewis:2022arg}, defined in Eq.~\ref{eq:dQ}, can be employed. The resulting \dq\ values, shown as open circles in Fig.~\ref{fig:fig2}, are consistent with those obtained directly from the $\pi$, $K$, and $p$ yields, demonstrating the reliability of the double ratio method. For completeness, \dq\ is also calculated using all charged particles via Eqs.~\ref{eq:Q} and \ref{eq:dQ_exact}, as shown by the open squares in Fig.~\ref{fig:fig2}. All three approaches yield nearly identical results.

\begin{figure*}[htb]
\centering{
\includegraphics[width=0.75\linewidth, angle=-0,keepaspectratio=true,clip=true]{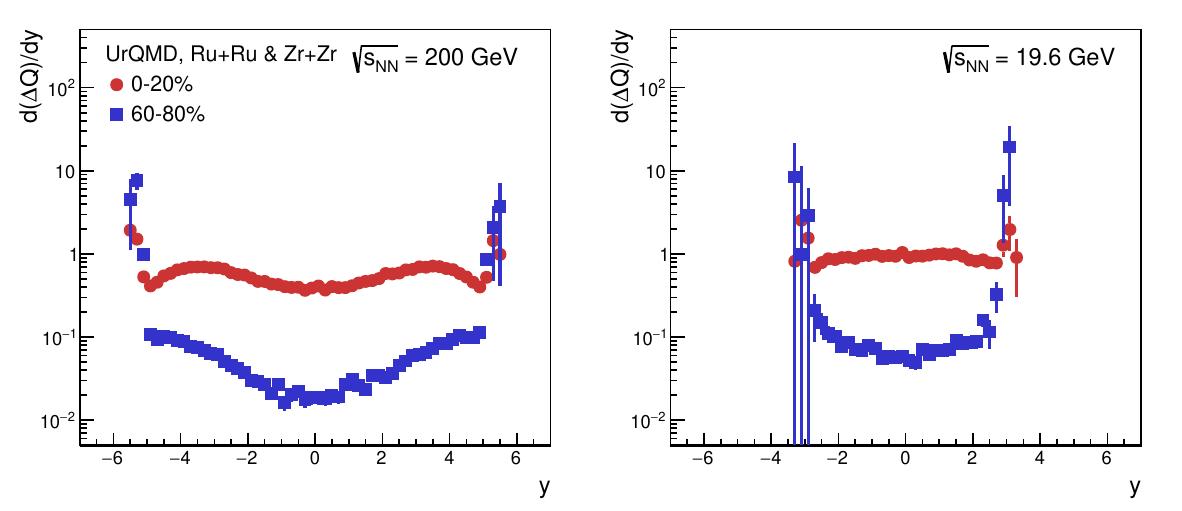}
\vskip -0.0cm
\caption{Rapidity distributions of $\Delta Q = Q_{\rm Ru+Ru} - Q_{\rm Zr+Zr}$ for 0--20\% and 60--80\% centrality classes at \roots\ = 200 (left) and 19.6 (right) GeV. 
\label{fig:fig3}
}
}
\vskip -0.0cm
\end{figure*}
The rapidity dependence of \dq\ is shown in Fig.~\ref{fig:fig3} for \roots\ = 200 and 19.6~GeV collisions. In 0--20\% central events, \dq\ remains relatively flat and enhanced around midrapidity, indicating substantial transport of electric charge from the beam region toward midrapidity. In contrast, for 60--80\% peripheral collisions, \dq\ is strongly suppressed relative to central events and becomes sharply peaked near the beam rapidity. This behavior suggests that, in peripheral (glancing) collisions, a large fraction of the initial-state charge is retained by the beam remnants.

\begin{figure*}[htb]
\centering{
\includegraphics[width=0.75\linewidth, angle=-0,keepaspectratio=true,clip=true]{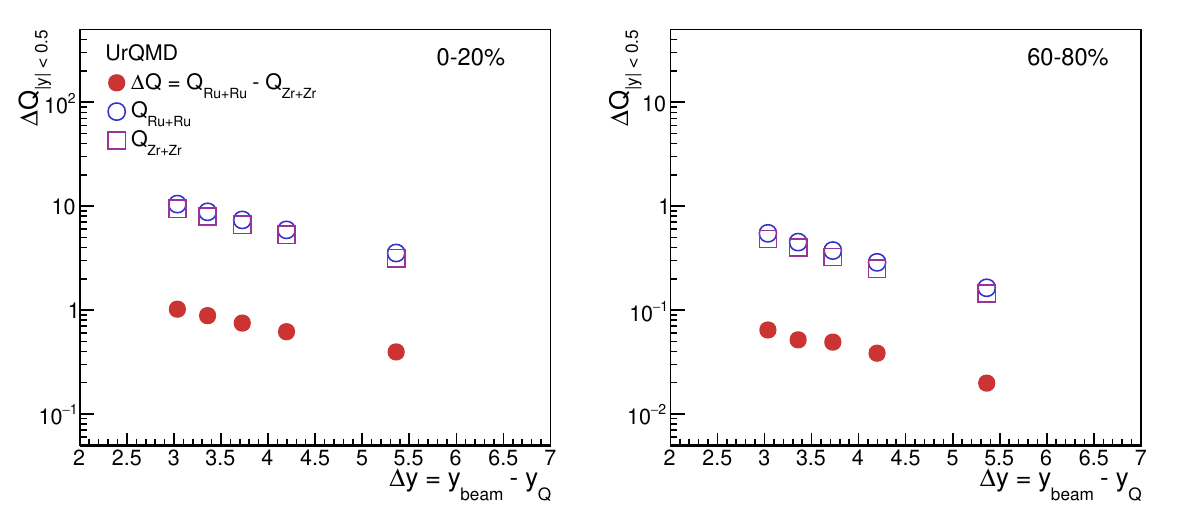}
\vskip -0.0cm
\caption{Distributions of $\Delta Q$ within $|y|<0.5$ against the rapidity gap ($\Delta y$), over which the electric charge is transported, for 0--20\% (left) and 60--80\% (right) Ru+Ru and Zr+Zr collisions simulated with UrQMD. Since $\Delta Q$ is evaluated around midrapidity, $y_{Q} = 0$. Similar distributions for $Q_{\rm Ru+Ru}$ and $Q_{\rm Zr+Zr}$ are also shown for comparison.
\label{fig:fig4}
}
}
\vskip -0.0cm
\end{figure*}

To characterize electric charge transport in heavy-ion collisions, we study \dq\ within $|y|<0.5$ as a function of the rapidity loss \dy, as shown in Fig.~\ref{fig:fig4} for central and peripheral collisions. The midrapidity yield of \dq\ exhibits an exponential decrease with increasing \dy, indicating that charge transport becomes less efficient over larger rapidity intervals. This behavior is qualitatively similar to that observed for baryon number transport~\cite{Lewis:2022arg}. For comparison, the electric charge yields in Ru+Ru and Zr+Zr collisions separately are also presented in Fig.~\ref{fig:fig4}. They follow a similar exponential trend but with magnitudes approximately one order of magnitude larger than that of \dq.

\begin{figure*}[htb]
\centering{
\includegraphics[width=0.85\linewidth, angle=-0,keepaspectratio=true,clip=true]{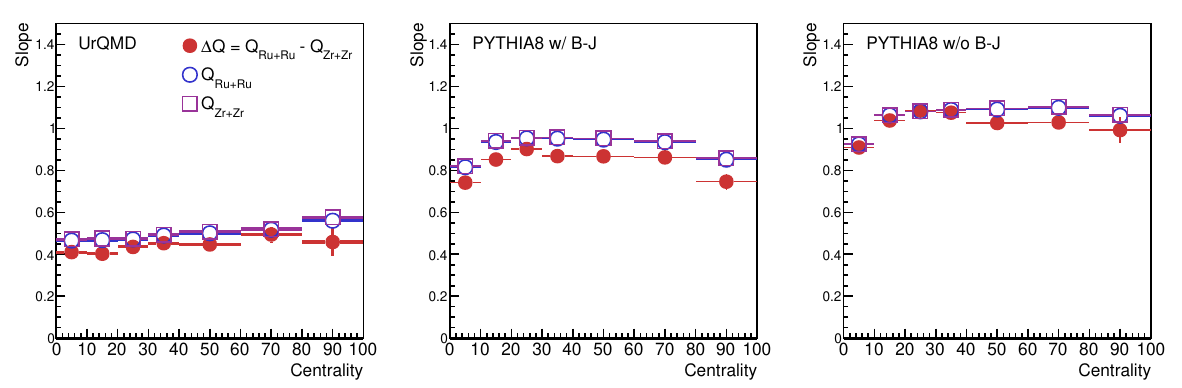}
\vskip -0.0cm
\caption{
Centrality dependence of the slopes for the $\Delta y$ dependence of $Q_{\rm Ru+Ru}$, $Q_{\rm Zr+Zr}$, and $\Delta Q$ within $|y|<0.5$ for Ru+Ru and Zr+Zr collisions. Different panels correspond to different event generators or generator configurations. Vertical bars around data points indicate fit errors on slopes.
\label{fig:fig5}
}
}
\vskip -0.0cm
\end{figure*}
The steepness of the decreasing trend is quantified by the slope parameter $\alpha_Q$, extracted by fitting the \dq\ distributions shown in Fig.~\ref{fig:fig4} with an exponential function, $C\times e^{-\alpha_Q \Delta y}$, where $C$ denotes a normalization constant. Figure~\ref{fig:fig5} presents the centrality dependence of the extracted $\alpha_Q$ values from UrQMD and the two \PYTHIA\ configurations. For UrQMD, a mild increase of $\alpha_Q$ from central to peripheral collisions is observed. This behavior may be attributed to enhanced multiple scatterings in central collisions, leading to stronger valence quark stopping and thus a smaller slope parameter. Such a centrality dependence contrasts with the experimentally observed centrality independence of the baryon transport slope $\alpha_B$~\cite{Lewis:2022arg, STAR:2024lvy}, indicating possibly different carriers for electric charge and baryon number. A similar increasing trend of $\alpha_Q$ is seen in \PYTHIA\ within the 0--30\% centrality interval, beyond which $\alpha_Q$ tends to saturate or slightly decreases. However, both \PYTHIA\ configurations, with and without baryon junctions, predict $\alpha_Q$ values about twice those predicted by UrQMD. This difference likely reflects genuine differences in how the two models describe heavy-ion collisions, rather than differences in the baryon transport mechanism. The inclusion of baryon junctions during hadronization in \PYTHIA\ slightly reduces the extracted $\alpha_Q$ compared to the version without junctions, implying that the junction mechanism can also enhance long-range transport of electric charge across large rapidity intervals. For comparison, the slope parameters extracted separately for $Q$ in Ru+Ru and Zr+Zr collisions are also shown in Fig.~\ref{fig:fig5}. They exhibit a similar centrality dependence to that observed for \dq, with slightly larger magnitudes. This consistency supports the robustness of using \dq\ as a probe of electric charge transport.

\begin{figure}[!h] 
\centering{
\includegraphics[width=0.9\linewidth, angle=-0,keepaspectratio=true,clip=true]{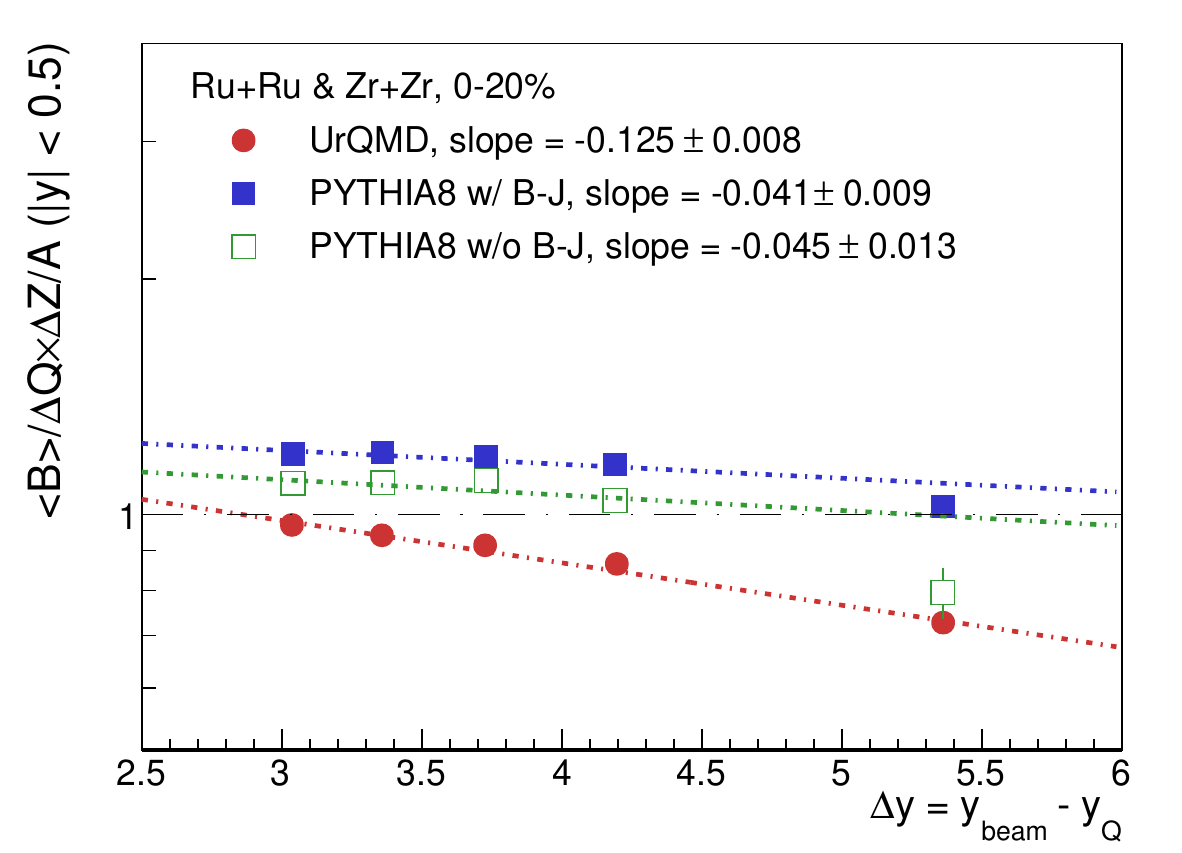}
\vskip -0.0cm
\caption{$\Delta y$ dependence of $\langle B\rangle/\Delta Q\times\Delta Z/A$ within $|y|<0.5$ for 0--20\% central Ru+Ru and Zr+Zr collisions. Different markers correspond to different model calculations, and dash-dotted lines indicate fits to data points with exponential functions. Slopes of these distributions from the fit are listed as well, along with statistical fit errors. 
\label{fig:fig6}
}
}
\vskip -0.0cm
\end{figure}
To directly compare the transport dynamics of electric charge and baryon number, we examine the ratio
\begin{equation}
R(\text{Isobar}) = \frac{\langle B\rangle}{\Delta Q} \times \frac{\Delta Z}{A},
\end{equation}
where $\langle B\rangle$ is the average baryon number between two isobaric systems, and $\Delta Z$ is the atomic number difference for the isobar nuclei. If baryon number is transported by the same valence quarks that carry electric charge, $R(\text{Isobar})$ is naively expected to be at unity when ignoring collision dynamics. Figure~\ref{fig:fig6} presents $R(\text{Isobar})$ at midrapidity as a function of \dy\ for 0--20\% central collisions from UrQMD and \PYTHIA, together with exponential fits. The UrQMD model exhibits a clear negative slope ($-0.125 \pm 0.008$). As \dy\ increases, its predicted $R(\text{Isobar})$ decreases below unity, which could arise from light quark stopping dynamics as observed in the AMPT model~\cite{Ross:2025qxr}.  \PYTHIA\ simulations with and without dynamical formation of baryon junctions during hadronization show similar but significantly shallower slopes compared to UrQMD. Moreover, the configuration including baryon junctions consistently yields larger values of $R(\text{Isobar})$ than the version without junctions, and maintains $R(\text{Isobar}) > 1$ from 19.6 to 200~GeV. This persistent excess above unity supports the interpretation that the inclusion of gluonic junctions, even only during the hadronization stage, can enhance baryon number transport over large rapidity intervals. However, measurements for 200~GeV Ru+Ru and Zr+Zr collisions by the STAR experiment at RHIC yield a $R(\text{Isobar})$ value significantly larger than those predicted by UrQMD and \PYTHIA\ at the same collision energy~\cite{STAR:2024lvy}. Furthermore, the observed decrease of $R(\text{Isobar})$ with increasing $\Delta y$ in both UrQMD and \PYTHIA\ is at odds with expectations from the genuine baryon-junction mechanism, in which baryon junctions are present in the nuclei and participate in the scattering process. Specifically, since baryon junctions are composed of low-momentum gluons, they are expected to be more readily stopped than high-momentum valence quarks~\cite{Kharzeev:1996sq}, which would lead to an increasing $R(\text{Isobar})$ with $\Delta y$. 

\begin{figure*}[!h]
\centering{
\includegraphics[width=0.80 \linewidth, angle=-0,keepaspectratio=true,clip=true]{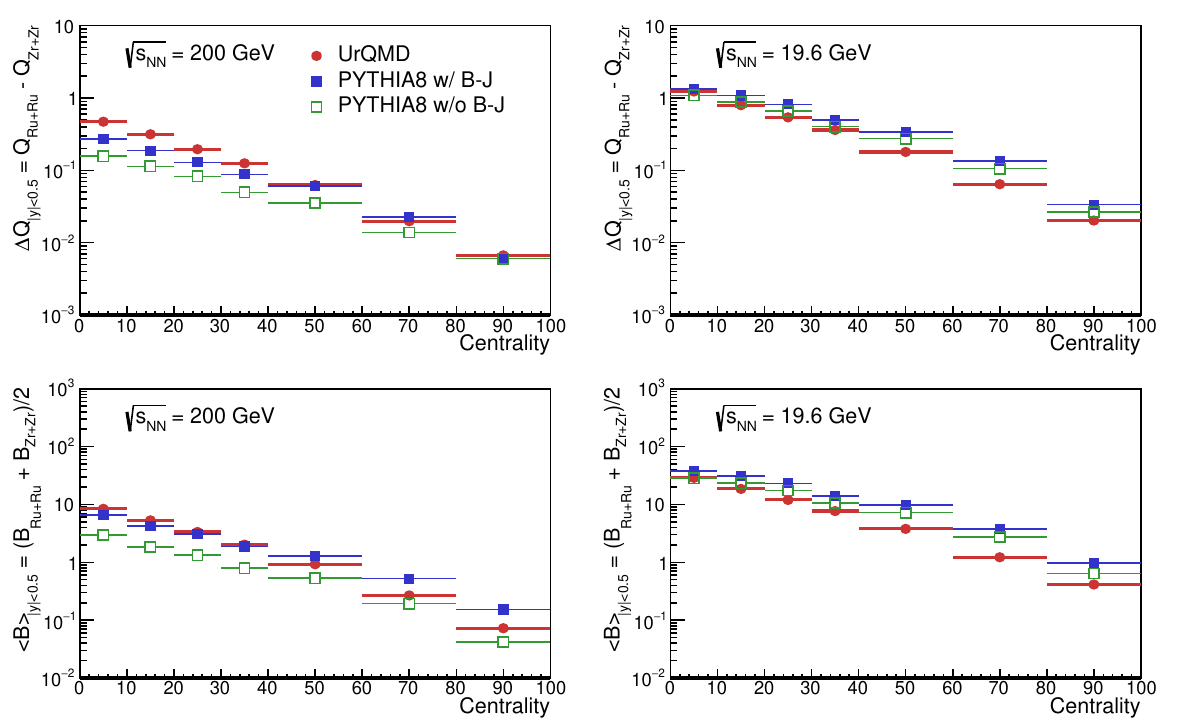}
\vskip -0.0cm
\caption{Comparison of $\Delta Q$ (top) and $\langle B\rangle$ (bottom) within $|y|<0.5$ as a function of centrality between UrQMD, \PYTHIA\ with B-J and \PYTHIA\ without B-J for Ru+Ru and Zr+Zr collisions at \roots\ = 200 (left) and 19.6 (right) GeV.
\label{fig:fig7} 
}
}
\vskip -0.0cm
\end{figure*}
To elucidate the different behaviors of $R(\text{Isobar})$ predicted by UrQMD and \PYTHIA, the centrality dependence of \dq\ and $\langle B\rangle$ within $|y|<0.5$ is shown in Fig.~\ref{fig:fig7} for Ru+Ru and Zr+Zr collisions at \roots\ = 200 and 19.6~GeV. At both energies and across all centrality classes, the inclusion of baryon junctions during hadronization in \PYTHIA\ leads to enhanced transport of electric charge and baryon number to midrapidity. At 200~GeV, UrQMD results are comparable to \PYTHIA\ with junctions in central collisions, while in peripheral events they are closer to the \PYTHIA\ configuration without junctions, for both electric charge and baryon transport. In contrast, at 19.6~GeV, UrQMD predicts systematically smaller transport of electric charge and baryon number to midrapidity than either \PYTHIA\ configuration in non-central collisions. Future measurements of electric charge and baryon transport in a beam energy scan of isobar collisions will be essential for discriminating between these model scenarios and for providing deeper insight into the underlying transport mechanisms.

\section{CONCLUSION}\label{sec:4}
In this work, we have presented a comprehensive study of electric-charge transport in high-energy nuclear collisions using isobaric Ru+Ru and Zr+Zr systems over a collision energy range of \roots\ = 19.6--200~GeV, based on simulations from the UrQMD and \PYTHIA\ event generators. By employing a double-ratio method, \dq\ between the two isobar systems can be determined with high precision. A beam-energy scan further enables \dq\ to be mapped as a function of \dy, thereby providing direct sensitivity to the dynamics of electric-charge transport.

We observe that \dq\ at midrapidity decreases exponentially with increasing \dy, indicating that electric-charge transport becomes progressively less efficient over larger rapidity intervals. The extracted slope parameter $\alpha_Q$ is smallest in central collisions, consistent with stronger multiple scatterings enhancing valence-quark stopping. Quantitatively, UrQMD predicts $\alpha_Q$ values roughly a factor of two smaller than those from \PYTHIA, while the inclusion of baryon junctions during hadronization in \PYTHIA\ modestly enhances long-range charge transport. A common feature across these models is that the rapidity slope of $R(\text{Isobar}) = \langle B\rangle/\Delta Q \times \Delta Z/A$ is always negative, reflecting a larger rapidity slope for baryon-number transport than for electric-charge transport. Notably, this behavior is opposite to expectations of a flatter baryon-number distribution than that of electric charge from baryon‑junction theory.

Overall, an energy scan of isobar collisions provides a realistic and experimentally feasible approach to precisely measuring electric-charge transport. Future measurements at the Electron-Ion Collider (EIC) and in the LHC fixed-target program will provide additional constraints on charge-redistribution models, help clarify the microscopic carriers of baryon number, and advance our understanding of conserved-charge transport in QCD matter.

\section*{Acknowledgement}
We thank the STAR Collaboration for supporting the inclusion of this proposal in its final Beam Use Request in 2026, prior to the permanent shutdown of RHIC.  This work is supported in part by National Natural Science Foundation of China under grant No. 12361141827 and the U.S. DOE Office of Science under contract Nos. DE-FG02-89ER40531, DE-SC0012704.




\bibliographystyle{elsarticle-num}
\bibliography{references} 

\end{document}